# Giant spin-orbit torque efficiency in all-epitaxial heterostructures

Nilamani Behera,[1,†] Rahul Gupta,[1] Sajid Husain[1,#], Jitendra Saha[2], Rajasekhar Pothala[2], Vineet Barwal,[3] Vireshwar Mishra[3], Gabriella Andersson[2], Dinesh K. Pandya,[3] Sujeet Chaudhary,[4] Rimantas Brucas,[1] Peter Svedlindh,[1*] and Ankit Kumar[1,ϒ,*]

[1]*Department of Materials Science and Engineering, Uppsala University, Box 35, SE-751 03 Uppsala, Sweden*
[2]*Department of Physics and Astronomy, Uppsala University, Box 516, SE-751 20 Uppsala, Sweden*
[3]*Departments of Physics and Materials Engineering, Indian Institute of Technology Jammu, Jammu 181221, India*
[4]*Thin Film Laboratory, Department of Physics, Indian Institute of Technology Delhi, New Delhi 110016, India*

**A large anti-damping spin-obit torque (SOT) efficiency in magnetic heterostructures is a prerequisite to realize energy efficient spin torque based magnetic memories and logic devices. The efficiency can be characterized in terms of the spin-orbit fields generated by anti-damping torques when an electric current is passed through the non-magnetic layer. We report a giant spin-orbit field of 48.96 (27.50) mT at an applied current density of 1 MAcm$^{-2}$ in *β*-W interfaced Co$_{60}$Fe$_{40}$ (Ni$_{81}$Fe$_{19}$)/TiN epitaxial structures due to an anti-damping like torque, which results in a magnetization auto-oscillation current density as low as 1.68 (3.27) MAcm$^{-2}$. The spin-orbit field value increases with decrease of *β*-W layer thickness, which affirms that epitaxial surface states are responsible for the extraordinary large efficiency. SOT induced energy efficient in-plane magnetization switching in large 20 × 100 μm$^2$ structures has been demonstrated by Kerr microscopy and the findings are supported by results from micromagnetic simulations. The observed giant SOT efficiencies in the studied all-epitaxial heterostructures are comparable to values reported for topological insulators. These results confirm that by utilizing epitaxial material combinations an extraordinary large SOT efficiency can be achieved using semiconducting industry compatible 5d heavy metals, which provides immediate solutions for the realization of energy efficient spin-logic devices.**

Present address:
[†]*Physics Department, University of Gothenburg, 412 96 Gothenburg, Sweden.*
[#]*Unité Mixte de Physique, CNRS, Thales, Université Paris-Saclay, 91767 Palaiseau, France.*
[ϒ]*IMEC, Kapeldreef 75, 3001 Heverlee, Belgium.*

Pure spin current based spintronic devices have advantages over conventional microelectronic devices owing to low energy dissipation, fast switching, and high-speed data processing., and can be integrated with microelectronic semiconductor devices for better functionality[1-4]. These new spintronic devices mainly work on the principle of spin manipulation employing the spin Hall effect (SHE) and the Rashba-Edelstein effect (REE). The basic building block of the spin devices is comprised of ferromagnetic (FM)/ non-magnetic (NM) bilayers; and the NM layers and their interfaces should possess strong relativistic spin-orbit interaction (SOI). The SOI can be of bulk as well as of interfacial origin, generating damping-like (DL) and field-like (FL) spin-orbit torques (SOTs)[4-8]. A charge current applied to SHE and REE based devices generates a transverse spin current and therefore a SOT at the FM/NM interface, which can be used to manipulate the FM state[4-7]. In contrast, devices based on the inverse effects, the inverse SHE (ISHE) and inverse REE (IREE) convert a spin current generated by spin pumping into a charge current in the NM layer by the ISHE and at the FM/NM interface by the IREE[6-12]. Bulk SOI of the NM layer is responsible for the SHE and ISHE mechanisms[2, 4, 5, 8], while interfacial SOI is responsible for the REE and IREE mechanisms[6, 7, 9, 10]. A strong DL SOT in FM (Ga,Mn)As thin films[7] and Py/CuO$_x$ heterostructures[10] has been reported, the origin of which lies in crystal inversion asymmetry and interfacial SOI induced Berry curvature[7, 10], respectively. High values of the spin Hall angle, which is a measure of the SHE, have been reported, of the order of 100% in Ni$_{0.6}$Cu$_{0.4}$ [13], 1880 % in the topological insulator Bi$_x$Se$_{1-x}$ [14], and 5200% in the conducting topological insulator Bi$_{0.9}$Sb$_{0.1}$ [15] thin films. The topological insulators exhibit the highest values of the spin Hall angle, however their thin film fabrication and integration in embedded memories are challenging. Therefore, present technological focus is to find CMOS (complementary metal-oxide semiconductor) technology compatible material combinations with even larger SOTs, or more specifically larger spin angular momentum transfer across the interfaces in magnetic heterostructures. Spin manipulation in the bulk or at the interface plays a decisive role in building the next generation of spintronic devices, viz. spin torque magnetic random-access memories (ST-MRAMs), spin logic devices, ST-transistors and ST nano-oscillators [4-10].

To optimize the bilayer stacks for devices there is a need to control interfacial properties like spin backflow, spin memory loss and magnetic proximity induced effects that may arise at/near the interface in FM/NM bilayer structures, properties that demean the overall spin transport in such structures[3, 16-22]. It is known that the spin current at the interface also exhibits spin memory loss in the presence of interfacial SOI, by creating a parallel relaxation path for the spin current[3, 19-22]. The spin memory loss can be quantified as the relative amount of spin current which is dissipated while passing through the FM/NM interface. Recently it was reported that interface alloying and abrupt interfaces may enhance the spin memory loss[22]. Therefore, it is apparent that to reduce the spin memory loss epitaxial interfaces constitute a potential solution.

In a quest to enhance the SOT efficiency of the 5d metal heterostructures, in this work we have fabricated $\beta$-W/Co$_{60}$Fe$_{40}$, Ni$_{81}$Fe$_{19}$/TiN/Si *all- epitaxial* heterostructures in which all the layers are epitaxial. The SOT efficiency of these structures has been determined by performing spin torque ferromagnetic resonance (STFMR) and planar Hall effect (PHE) measurements. In-plane magnetization switching has been evidenced by recording Kerr microscopy images. The dc bias dependent changes of the effective damping and the PHE results confirm the presence of significantly large interfacial DL torques in these heterostructures whose origin lies both at the $\beta$-W and TiN interfaces with the epitaxial ferromagnetic layer. The SOT induced magnetization auto-oscillations and magnetization switching current density at which the effective damping reverses sign is comparable to the values reported for conductive topological insulators[15].

**Structural Characterization** To confirm the epitaxial growth of CoFe and NiFe (henceforth referred to as Py) on the *epi*-TiN buffered Si substrate, texture analyses were performed by recording X-ray diffraction (XRD) pole figures and scanning transmission electron microscopy (STEM) images.

Figure 1 (a) shows the pole figure XRD patterns of the CoFe(022) plane at $2\theta=45.2^o$ for the W/CoFe/TiN/Si thin films, while Fig. 1(b) shows the pole figure XRD patterns of the Py(111) plane at $2\theta = 44.2^o$ for the W/Py/TiN/Si thin films confirming the epitaxial quality of the thin films; here W is read as β-W. Figures 1(c) and (d) show the cross-section STEM images of the W/CoFe/TiN/Si and W/Py/TiN/Si heterostructures, respectively. The STEM images give evidence of sharp epitaxial interfaces in both the structures. The interface roughness of each individual layer determined by X-ray reflectivity is in the range of ≤ 1 nm, and these values are also closely matching with previously reported values[23-25].

**Spin-orbit torque ferromagnetic resonance** To determine the SOTs, STFMR measurements were performed on patterned (20×100 μm$^2$) *all-epitaxial* W/CoFe, Py/TiN/Si structures. Schematic figures of the STFMR setup and the torques acting on the magnetization are shown in Figs. 2(a) and (b) (see Ref. 26 for measurement details). In the STFMR measurements, the microwave current $I_{rf}$ of a constant frequency $f$ generates an Oersted field and a transverse spin current in the NM layer and at its interface with the FM layer. The torques due to the Oersted field ($\tau_{Oe}$) and due to the transverse spin current ($\tau_{AD}$) contribute with anti-symmetric and symmetric profiles, respectively, to the FMR line-shape. The anti-damping $\tau_{AD}$ torque, dominantly generated by the transverse spin current, acts against the intrinsic damping torque $\tau_D$ in the FM layer. In case of interfacial SOI, the REE contributes dominantly with a field-like torque ($\tau_{FL}$) with direction opposite to that of $\tau_{Oe}$. At resonance, the temporal variation of the magnetization vector with respect to the direction of $I_{rf}$, in combination with the anisotropic magnetoresistance of the FM layer and spin magnetoresistance of the interface, generates a time varying resistance that mixes with $I_{rf}$ yielding a dc voltage output. In our case, using a low-frequency (1 kHz) modulation of $I_{rf}$ the STFMR signal is detected using a lock-in amplifier. The observed STFMR spectra exhibit a combination of symmetric and anti-symmetric Lorentzian weight factors[27-33], as shown in Figs. 2(c) and (d). The STFMR spectrum at constant $f$ is expressed as $V_{mix} = V_0 [SF_S(H) + AF_A(H)]$, where $V_0$ is the amplitude of the mixing voltage. $S$ and $A$ are symmetric and anti-symmetric Lorentzian weight factors, accounting for anti-damping and field-like torques, respectively. $F_S(H)$ is the symmetric and $F_A(H)$ is the anti-symmetric Lorentzian function (for details see Ref. 28).

In order to make a reliable determination of the SOT efficiency, one can use the dc induced change of the damping, often referred to as the modulation of damping (MOD) method, for which the spin pumping (ISHE) and field-like contributions are absent. In the MOD method, one measures the change of the STFMR linewidth ($\Delta H$) applying a constant dc to the patterned structure. From the linewidth versus frequency behavior we can determine how the effective damping value varies with the applied dc [27-29].

The $I_{dc}$ induced change of the effective Gilbert damping $\alpha_{eff}(I_{dc})$ is given by[28,29]

$$\alpha_{eff}(I_{dc}) - \alpha_{eff}(I_{dc} = 0) = \left(\frac{\sin\varphi}{(H_r + 0.5 M_{eff})\mu_0 M_S t_{FM}} \frac{\hbar}{2e}\right) J_S, \quad (1)$$

where $\varphi$, $H_r$, $M_{eff}$, $M_S$, $\hbar$ and $e$ are the angle between the current and the magnetization, resonance field, effective magnetization, saturation magnetization, reduced Planck's constant and electric charge of an electron, respectively. The spin current density, $J_S = \theta_{SH}^{MOD} J_{C,dc}(W + TiN)$;

$$J_{C,dc}(W + TiN) = I_{dc}\left[\left(\frac{1}{A_W}\frac{R_{FM} \times R_{TiN}}{(R_W \times R_{TiN}) + (R_{FM} \times R_{TiN}) + (R_W \times R_{FM})}\right) + \left(\frac{1}{A_{TiN}}\frac{R_{FM} \times R_W}{(R_W \times R_{TiN}) + (R_{FM} \times R_{TiN}) + (R_W \times R_{FM})}\right)\right].$$

Here $J_{C,dc}$ is the charge current density in the non-magnetic layers. $A_W$ and $A_{TiN}$ are the cross-sectional areas of the W and TiN layers, and $R_W$, $R_{TiN}$ and $R_{FM}$ are the resistances of the W, TiN and FM layers, respectively. The measured resistivity values of W, TiN, CoFe and Py are 3 µΩ-cm, 38 µΩ-cm, 32 µΩ-cm, and 31 µΩ-cm, respectively. The low resistivity of W is due to the presence of a conductive native $WO_x$ in $\beta$-W. In the W(6nm)/CoFe(10nm)/TiN/Si structure 88% of $I_{dc}$ passes through the W and TiN/Si layers, while in the W(8nm)/Py(15nm)/TiN/Si structure 84% of $I_{dc}$ passes through the W and TiN/Si layers. The spin Hall angle $\theta_{SH}^{MOD}$ can be estimated by measuring the $I_{dc}$ dependent rate of change of the effective damping, $\partial \alpha_{eff}(I_{dc})/\partial I_{dc}$. Hence, $\theta_{SH}^{MOD}$ can be expressed as[29]

$$\theta_{SH}^{MOD} = \left[\frac{\partial \alpha_{eff}/\partial J_{C,dc}(W+TiN)}{\left(\frac{\sin\varphi}{(H_r+M_{eff})\mu_0 M_S t_{FM}}\frac{\hbar}{2e}\right)}\right]. \quad (2)$$

STFMR spectra were recorded for different $I_{dc}$ in the range +5 mA to −5mA on *all-epitaxial* W(6nm)/CoFe(10nm)/TiN/Si and W(8nm)/Py(15nm)/TiN/Si. The recorded spectra were fitted by using Lorentzian functions to determine the line-shape parameters. Figure 2(e) shows $f$ vs. $H_r$ data and figure 2(f) shows $\Delta H$ vs. $f$ data at respective applied dc for the two multi-layer structures. The $f$ vs. $H_r$ data was fitted by using the in-plane Kittel equation to determine the gyromagnetic ratio and $M_{eff}$. The $\Delta H$ vs. $f$ data was fitted by using the standard expression to determine $\alpha_{eff}(I_{dc})$, which are presented in Fig. 3(a). $\alpha_{eff}(I_{dc})$ increases from 0.0025±0.0006 to 0.0117±0.0004 for W(6nm)/ CoFe(10nm)/TiN/Si as $I_{dc}$ is varied from +5mA to −5mA. To make this data more readable, the change in $\Delta H$ and $\alpha_{eff}$ with $I_{dc}$, i.e. $\Delta H(I_{dc}) - \Delta H(I_{dc} = 0)$ and $\alpha_{eff}(I_{dc}) - \alpha_{eff}(I_{dc} = 0)$ vs. $I_{dc}$, are plotted in Figs. 3(b) and (c). A linear decrease in $\Delta H(I_{dc}) - \Delta H(I_{dc} = 0)$ with increasing $I_{dc}$ for both samples clearly depicts the absence of heating in our measurements. The observed variation of $\alpha_{eff}(I_{dc})$ is larger in W(6nm)/CoFe(10nm)/TiN/Si in comparison to W(8nm)/Py(15nm)/TiN/Si,. The percentage current-induced change of $\alpha_{eff}(I_{dc})$, defined as, $\frac{\alpha_{eff}(I_{dc}) - \alpha_{eff}(I_{dc}=0)}{\alpha_{eff}(I_{dc}=0)} \times 100\%$, at $I_{dc} = \pm 5$ mA is 66% and 26% for the W based CoFe and Py heterostructures, respectively. Here, $I_{dc} = 5mA$ corresponds to a dc current density $J_{dc} = 1.07 \times 10^{10} \frac{A}{m^2}$ and $6.61 \times 10^9 \frac{A}{m^2}$ in the W layer for the CoFe and Py based heterostructures, respectively. The observed current-induced change of $\alpha_{eff}(I_{dc})$, 6.14% (3.89%), in *all-epitaxial* W/CoFe,Py/TiN/Si is significantly larger than that reported by Pai *et al.*[31], where a current-induced change of 0.17% of the effective damping constant at $J_{dc} = \pm 1.0 \times 10^9 \frac{A}{m^2}$ was reported for the W(6nm)/Co$_{40}$Fe$_{40}$B$_{20}$(5nm)/SiO$_x$/Si system. Our current-induced change of $\alpha_{eff}(I_{dc})$ is also significantly larger than that reported by Liu *et al.*[27], who achieved a current-induced change of $\alpha_{eff}(I_{dc})$ of 0.003% at $J_{dc} = \pm 8.95 \times 10^9 \frac{A}{m^2}$ in Py(4nm)/Pt(6nm) structures. The calculated value of $|\theta_{SH}^{MOD}|$ averaged over all measured frequencies for the W/CoFe (Py)/TiN/Si heterostructure using Eq. (4) is 39.01±0.84 (8.45±0.27). The SOT efficiency from the TiN/CoFe(Py) interface is 1.22±0.47 (0.25±0.19). The critical switching current density of the *all-epi* W/CoFe (Py)/TiN/Si heterostructure at which effective damping reverses sign is 1.68 (3.27) MA/cm$^2$, while it is 5.99 (9.76) MA/cm$^2$ for a heterostructure not including the TiN buffer layer. The SOT efficiency determined using Eq. (2) depends on the thickness and magnetization of the magnetic layer. The spin-orbit field, which is generated when feeding a charge current through the NM layer due to the SHE or REE, is independent of magnetic layer thickness and magnetization. By measuring spin-orbit field per unit applied current density, one can predict the real feature of the heterostructures .

**Angle dependent planar Hall effect measurements to determine spin-orbit torques** To evaluate the SOT efficiency we have determined the out-of-plane spin-orbit field in W/CoFe, Py/TiN/Si heterostructures by performing angle dependent planar Hall effect (PHE) measurements. An optical image of the Hall bar structure is shown in Fig. 4 (a), where an in-plane dc is applied along the length of the bar and the voltage is recorded across the bar while varying the in-plane magnetic field direction. The vector representation of the spin-orbit field in the PHE measurements is presented in Fig. 4 (b), where $\varphi$ is the angle between the current direction and the in-plane applied field $H$ and $\theta$ is the angle between the current direction and the magnetization vector $M$. The rotation angle $\phi$ is considered here with the initial alignment $\phi = \pi/2$, yielding $\varphi = \phi - \pi/2$. The angle dependent planar Hall resistance recorded at opposite current polarities can be expressed as[14]

$$R(\pm I_{dc}, \varphi) = R_0 + R_{AHE} \cos \varphi + R_{PHE} \sin 2\varphi, \qquad (3)$$

where $I_{dc}$ is the applied dc, $R_0$ is the resistance offset accounting for the Hall bar imbalance, $R_{PHE}$ and $R_{AHE}$ are the resistances due to the PHE and anomalous Hall effect (AHE), respectively. The spin-orbit field $H_{SO}$ was determined from the differential Hall resistance $R_{DH}$ defined as[14],

$$R_{DH}(I_{dc}, \varphi) = R(I_{dc}, \varphi) - R(-I_{dc}, \varphi) = 2R_{PHE} \frac{H_{Oe} + H_{FL}}{H} (\cos \varphi + \cos 3\varphi) +$$

$$+2 \frac{dR_{AHE}}{dH_{op}} H_{SO} \cos \varphi + C, \qquad (4)$$

where $H_{Oe}$ is the Oersted field due to $I_{dc}$, $H_{FL}$ is the field due to the REE contributing with a field-like torque ($\tau_{FL}$), $\frac{dR_{AHE}}{dH_{op}}$ was obtained from a separate measurement of the AHE resistance $R_{AHE}$ in an out-plane applied magnetic field $H_{op}$ and $C$ is the resistance offset. The extracted values of $\frac{dR_{AHE}}{dH_{op}}$ are 0.026 Ω/T and 0.096 Ω/T for the CoFe and Py films, respectively (see Supplementary section S5 and ref#34). The $R_{DH}$ data was recorded at $H$ =0.5 T, which is large enough to saturate the magnetization of the FM (CoFe, Py) layers and to suppress the contribution from the field-like torque. Figures 4 (c) and (d) show the PHE resistance $R(I_{dc}, \varphi)$ versus $\varphi$ recorded with opposite current polarities for W/CoFe/TiN/Si and W/Py/TiN/Si, respectively. The experimental data were fitted using Eq. (3) to determine the values of $R_{PHE}$ and $R_{AHE}$. The variation of the spin-orbit field $H_{SO}$ with $I_{dc}$ was derived by fitting the angle dependent $R_{DH}(I, \varphi)$ data to Eq. (4) as shown in Figs. 4 (e) and (f) for the W(8nm)/CoFe(10nm)/TiN/Si and W(8nm)/Py(15nm)/TiN/Si structures, respectively. The results from this fitting are presented in Fig. 4 (g), showing $H_{SO}$ versus $J_{dc}$.

The PHE determined SOT values depend on thickness and magnetization of the ferromagnetic layer, therefore we should rely more on the $H_{SO}$ derived values to understand the heterostructure SOT efficiency. To dig deeper into the origin of these extraordinary large $H_{SO}$ values, we have investigated W-thickness ($t_W$) dependence of $H_{SO}$ in W(4-10nm)/Py(15nm)/TiN(15nm); the results are shown in Figs. 4 (h-k). We have not varied the TiN thickness here since the TiN/Py interface spin torque efficiency is very small compared to that of the Py(15nm)/W interface. The angle dependent planar Hall resistance at $I_{dc} = \pm 4.5$ mA dc is presented in Fig. 4 (h), while the differential Hall resistance is plotted in Fig. 4(i) for the W(4nm)/Py(15nm)/TiN(15nm)/Si structure for different dc amplitudes. Figure 4(j) shows $\mu_0 H_{SO}$ versus $J_{dc}$ for different $t_W$, while Fig. 4(k) shows $\mu_0 H_{SO}$ versus $t_W$ at constant $J_{dc} = 1 \times 10^{10}$ A/m². It is evident from the results that $\mu_0 H_{SO}$ at constant $J_{dc}$ increases with decreasing W-thickness.

**In-plane magnetization switching** To show SOT driven magnetization switching we have recorded MOKE microscope images on rectangular bar structures of size $20 \times 100$ $\mu m^2$ for W/CoFe(Py)/TiN/Si by varying the amplitude of the applied dc pulse. Since the device structure is large, possessing a multidomain magnetic microstructure in the remanent magnetization state, we have applied a magnetic field of 4.5 mT (0.35 mT) along the $\pm y$ directions for the W/CoFe(Py)/TiN/Si structures while applying the dc pulse along the $\pm x$ directions. Figure 5 (a) shows MOKE images recorded for the W(7nm)/Py(15nm)/TiN heterostructure. A black contrast in the MOKE image represents a domain magnetization along the $+y$-direction, while a white contrast represents a domain magnetization along the $-y$-direction. By applying the magnetic field along the $+y$-direction a mixed dark-light contrast appears in the MOKE images, which represents a multidomain magnetic state. The contrast gradually changes to dark by increasing the amplitude of the dc pulse along the $+x$-direction. To further confirm the SOT switching, the magnetic field was applied along the $-y$ direction and the pulsed dc was applied along the $-x$ direction. The change in magnetization contrast by changing the direction of the applied dc pulse confirms the in-plane magnetization switching. The applied current density value required for room temperature SOT induced in-plane magnetization switching was $J_{dc} = \pm 5.78 \times 10^{10}$ $\frac{A}{m^2}$ for the W(8nm)/Py(15nm)/TiN structure. The MOKE switching for CoFe is described in supplementary information S6. The absence of magnetization switching by changing the applied dc pulse direction while keeping the applied magnetic field the in $-y$-direction affirms the absence of Joule heating or Oersted field induced magnetization switching.

SOT induced magnetization switching was also studied in the remnant magnetization state with the result that only a small part of the magnetic microstructure could be switched (see supplementary S6). The remanent magnetization state exhibits a multidomain magnetic microstructure and the SOT acting on the domain magnetizations will only switch domains with a domain magnetization along the hard-axis of the bar shaped structure ($\pm y$-directions). The SOT magnetization switching has also been modeled by using micromagnetic simulations on a rectangular cuboid structure model with dimensions $200 \times 1000 \times 15 \text{nm}^3$ (the same aspect ratio in-plane as used in the STFMR and Hall experiments). The experimentally determined layer parameters, i.e., effective damping constant, anti-damping and field-like torques, effective anisotropy, saturation magnetization and thicknesses, were used in the simulations. The dc pulse dependent magnetization switching is presented in Figs. 5 (b) and (c) for the W/Py/TiN structure. The simulated in-plane magnetization switching current density value was $J_{dc} = \pm 7.0 \times 10^{10} \frac{A}{m^2}$ and is comparable to the experimentally determined value.

**Discussion** The determined giant value of the spin-orbit field is comparable to values reported for conducting topological insulators; $Bi_{0.9}Sb_{0.1}$/MnGa bilayers[15] and $Bi_xSe_{(1-x)}$($t_{BS}$ nm)/CoFeB(5 nm)[14]. Therefore, in a quest to find the origin of the observed giant spin-orbit torque efficiency in *all-epi* heterostructures, we have delved into the literature on interfacial SOT and/or presence of topological surface states in W.

Thonig *et. al.*[35] have reported Dirac-type surface states in the (110) plane of strained W, which are topologically protected by mirror symmetry and, thus, exhibiting nonzero topological invariants. The presence of these topological states near the Fermi level can yield large spin-orbit torque efficiency. Very recently, theoretical calculations have also revealed that β-W is a topological massive Dirac metal[36]. In the absence of SOC, in calculations, Dirac nodal lines exist in the Brillouin zone and their existence is associated with the band inversion at high-symmetry points. However, once the SOC is switched-on in the calculations massive Dirac states appear. Furthermore, the nontrivial topologically protected surface states and Fermi arcs have also been observed[36] in single crystals. This work proposed that epitaxial β-tungsten is a pure topological massive Dirac metal. Both these theoretical reports suggest that both the epitaxial and strained W (110) plane can exhibit massive Dirac surface states. In our studied *all- epitaxial* systems β-W is also epitaxial and exhibits a strain with respect to the FM epitaxial interface. Therefore, a presence of the Dirac surface states at the β-W interface in our studied all-epitaxial heterostructures should be considered.

A giant spin torque efficiency was in the topological insulator $Bi_xSe_{1-x}$, which increased with decreasing film thickness owing to topological surface states induced momentum locking[14]. We have also observed an enhancement of the spin torque efficiency and the spin-orbit field by reducing the thickness of the β-W layer, which clearly evidences the presence of topological surface states. Therefore, the observed giant SOT efficiency and the spin-orbit field are due to the presence of surface topological states providing a large spin Berry curvature and hence a large intrinsic SHE in the *all-epitaxial* W based heterostructures[36,37]. Moreover, light element, e.g. Ti, interfaces with in-plane magnetized FM generates large interfacial spin-orbit torques as demonstrated in CoFeB/Ti, NiFe(4nm)/Ti(3nm)[38]. TiN also generates a non-negligible spin current at the FM/TiN interface and we find that the FM/TiN interface also plays a role in the $I_{dc}$ dependent change of the effective damping, originating from anti-damping-like torques due to interfacial symmetry breaking and SOI. The interfacial spin torques are very sensitive to the crystallographic structure and therefore to the orbital ordering at the interface. Hence, it appears that it is the epitaxial nature of the heterostructures and the topological surface states that are responsible for the observed giant spin-orbit torque efficiency.

**Conclusions**

In summary, spin-orbit torque studies have been performed on all-epitaxial β-W/CoFe/TiN/Si and β-W/Py/TiN/Si heterostructures. The observed giant spin torque efficiency is comparable to values reported for topological insulators, and the origin of this giant spin-orbit torque efficiency lies in the creation of topological surface states at the epitaxial interfaces. The extraordinary spin-orbit torque efficiency in the studied semiconducting technology compatible W and TiN interfaced FM heterostructures opens up a new avenue for the realization of ultralow powered spintronic devices by utilizing epitaxial magnetic heterostructures.

**Materials and Methods**

The epitaxial $Co_{60}Fe_{40}$ and $Ni_{81}Fe_{19}$(Py) films were deposited on TiN(200)[100](10nm)/Si substrates by pulsed dc magnetron sputtering technique. Layers of W with different thickness $t_W$ in the range 1-15 nm were deposited on CoFe/TiN/Si and Py/TiN/Si. The FM layer thickness $t_{FM}$ was varied in the range 3-17nm. It was noted that the grown W layers mostly exhibit the desired *β*-phase (i.e., A-15 cubic for W) as reported in our previous work[23,39]. X-ray reflectivity, X-ray diffraction pole figures and transmission electron microscopy measurements were performed to ascertain the quality of the films. Spin-torque measurements were performed using a home build ultralow noise STFMR setup. Angle dependent PHE measurements were performed in a Quantum Design Physical Property Measurement System (PPMS). Kerr images were recorded using an Evicomagnetics Kerr microscope system in longitudinal geometry. Micromagnetic simulations were performed by using the MuMax3 framework to solve the time and space-dependent Landau-Lifshitz-Gilbert-Slonczewski equation for the magnetization. The details of the experimental methods are presented in the SI file.

**References:**

1. W. Zhang, W. Han, X. Jiang, S.-H. Yang, and S. S. P. Parkin, Nat. Phys. 11, 496–502 (2015).
2. K. Ando, T. Yoshino, and E. Saitoh, Appl. Phys. Lett. 94, 152509 (2009).
3. J.-C. Rojas-Sánchez, N. Reyren, P. Laczkowski, W. Savero, J.-P. Attané, C. Deranlot, M. Jamet, J.-M. George, L. Vila, and H. Jaffrès, Phys. Rev. Lett. 112, 106602 (2014).
4. A. Brataas, A. D. Kent and H. Ohno, Nat. Mater. 11, 372 (2012)
5. K. Ando, S. Takahashi, K. Harii, K. Sasage, J. Ieda, S. Maekawa, and E. Saitoh, Phys. Rev. Lett. 101, 036601 (2008).
6. I. M. Miron, G. Gaudin, S. Auffret, B. Rodmacq, A. Schuhl, S. Pizzini, J. Voge, and P.Gambardella, Nat. Mater. 9, 230–234 (2010).
7. H. Kurebayashi, J. Sinova, D. Fang, A. C. Irvine, T. D. Skinner, J. Wunderlich, V. Novák, R. P. Campion, B. L. Gallagher, E. K. Vehstedt, L. P. Zârbo, K. Výborný, A. J. Ferguson and T. Jungwirth, Nat. Nanotechnol. 9, 211 (2014).
8. E. Saitoh, M. Ueda, H. Miyajima, and G. Tatara, Appl. Phys. Lett. 88, 182509 (2006).
9. J. C. Rojas-Sánchez, L. Vila, G. Desfonds, S. Gambarelli, J.P. Attané, J. M. D Teresa, C. Magén and A. Fert, Nat. Commun. 4, 2944 (2013).
10. T. Gao, A. Qaiumzadeh, H. An, A. Musha, Y. Kageyama, J. Shi, and K. Ando, Phys. Rev. Lett. 121, 017202 (2018).
11. O. Mosendz, J. E. Pearson, F. Y. Fradin, G. E. W. Bauer, S. D. Bader, and A. Hoffmann, Phys. Rev. Lett. 104, 046601 (2010).
12. H. Nakayama, K. Ando, K. Harii, T. Yoshino, R. Takahashi, Y. Kajiwara, K. Uchida, and Y. Fujikawa, Phys. Rev. B 85, 144408 (2012).
13. M. W. Keller, K. S. Gerace, M. Arora, E. K. Delczeg-Czirjak, J. M. Shaw, and T. J. Silva, Phys. Rev. B 99, 214411 (2019)
14. Mahendra DC, R. Grassi, J. Y. Chen, M. Jamali, D. R. Hickey, D. Zhang, Z. Zhao, H. Li,

P. Quarterman, Y. Lv, M. Li, A. Manchon, K. A. Mkhoyan, T. Low & J. P. Wang, Nat. Mat., 17, 800–807 (2018).

15. N. H. D. Khang, Y. Ueda & P. N. Hai, Nat. Mat. 17, 808–813 (2018).
16. Y. Tserkovnyak, A Brataas, G. E. W. Bauer, and B. I. Halperin, Rev. Mod. Phys. 77, 1375–1421 (2005).
17. A. Brataas, Y. Tserkovnyak, G.E.W. Bauer, and B.I. Halperin, Phys. Rev. B 66, 060404(R) (2002).
18. H. Jiao and G. E. W. Bauer, Phys. Rev. Lett. 110, 217602 (2013).
19. K. Chen and S. Zhang, Phys. Rev. Lett. 114, 126602 (2015).
20. K. Chen and S. Zhang, IEEE Magn. Lett. 6,1 –4 (2015).
21. H. Kurt, R. Loloee, W.P. Pratt, Jr., and J. Bass, Appl. Phys. Lett. 81, 4787 (2002).
22. K. Gupta, R. J. H. Wesselink, R. Liu, Z. Yuan, and P. J. Kelly, Phys. Rev. Lett. 124, 087702 (2020).
23. N. Behera, P. Guha, D. K. Pandya, and S. Chaudhary, ACS Appl. Mater. Interfaces 9, 31005−31017 (2017).
24. R. Bansal, N. Behera, A. Kumar, and P. K. Muduli, Appl. Phys. Lett. 110, 202402 (2017).
25. N. Behera, S. Chaudhary, and D. K. Pandya, Sci. Rep. 6, 19488 (2016).
26. A. Kumar, S. Akansel, H. Stopfel, M. Fazlali, J. Akerman, R. Brucas, and P. Svedlindh, Phys. Rev. B 95, 064406 (2017).
27. L. Liu, T Moriyama, D. C. Ralph, and R. A. Buhrman, Phys. Rev. Lett. 106, 036601 (2011).
28. S. Kasai, K. Kondou, H. Sukegawa, S. Mitani, K. Tsukagoshi, and Y. Otani, Appl. Phys. Lett. 104, 092408 (2014).
29. K. U. Demasius, T. Phung, W. Zhang, B. P. Hughes, S.-H. Yang, A. Kellock, W. Han, A. Pushp, and S. S. P. Parkin, Nat. Commun. 7, 10644 (2016),
30. A. Ganguly, K. Kondou, H. Sukegawa, S. Mitani, S. Kasai, Y. Niimi, Y. Otani, and A. Barman, Appl. Phys. Lett. 104, 072405 (2014).
31. C. F. Pai, L. Liu, Y. Li, H. W. Tseng, D. C. Ralph, and R. A. Buhrman, Appl. Phys. Lett. 101, 122404 (2012).
32. L. Liu, C. F. Pai, Y. Li, H. W. Tseng, D. C. Ralph, and R. A. Buhrman, Science 336, 555 (2012).

33. G. Allen, S. Manipatruni, D. E. Nikonov, M. Doczy, and I. A. Young, Phys. Rev. B 91, 144412 (2015).

34. S. Husain, X. Chen, R. Gupta, N. Behera, P. Kumar, T. Edvinsson, F. García-Sánchez, R. Brucas, S. Chaudhary, B. Sanyal, P. Svedlindh, and Ankit Kumar, Nano Lett. 20, 6372−6380, (2020)

35. D. Thonig, T. Rauch, H. Mirhosseini, J. Henk, I. Mertig, H. Wortelen, B. Engelkamp, A. B. Schmidt, and M. Donath, Phys. Rev. B 94, 155132 (2016).

36. J. Li, S. Ullah, R. Li, M. Liu, H. Cao, D. Li, Y. Li, and X-Qiu Chen, Phys. Rev. B 99, 165110 (2019).

37. X. Sui, C. Wang, J. Kim, J. Wang, S. H. Rhim, W. Duan, N. Kioussis, Phys. Rev. B 96, 241105(R) (2017).

38. S. C. Baek, V. P. Amin, Y. W. Oh, G. Go, S. J. Lee, G. H. Lee, K. J. Kim, M. D. Stiles, B. G. Park and K. J. Lee, Nat. Mater. 17, 509 (2018).

39. N. Behera, A. Kumar, S. Chaudhary, and D. K. Pandya, RSC Adv. 7, 8106 (2017).
**Acknowledgements**

This work was supported by the Swedish Research Council (VR), grant no 2017-03799. Prof. Mikael Ottosson is acknowledged for help during XRD measurements.
**Authors Contributions:**

NB fabricated all films with a support from AK, SC, and DP. NB performed STFMR with support from AK. NB performed XRR measurements. NB performed pole figure measurements. S.H and R.G performed Hall measurements. SH performed micromagnetic simulations. RB fabricated all the devices. JS and RP performed Kerr microscopic measurements with support from GA. NB, SH, PS and AK analysed the data. NB and AK wrote the manuscript. AK conceived the idea. AK and PS designed and supervised this project. All authors discussed the results, reviewed and commented on the manuscript.

*Correspondence and request for materials should be addressed to Email:

chainutyagi@gmail.com

peter.svedlindh@angstrom.uu.se
16

**Competing financial interest**

The authors declare no competing financial interests.

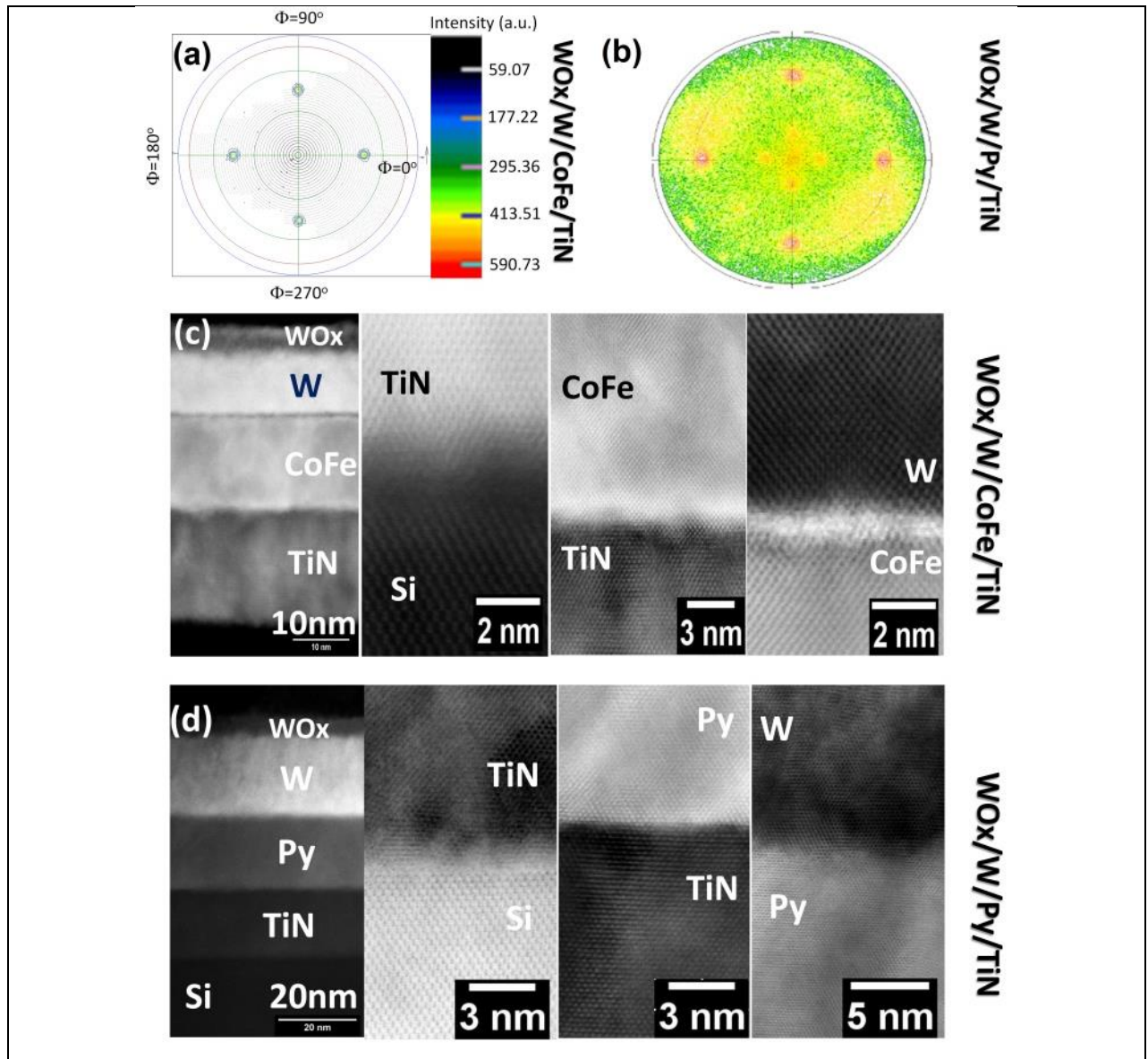

**Figure 1** Pole figure XRD patterns of (a) $Co_{60}Fe_{40}$ (022) in W/$Co_{60}Fe_{40}$/TiN/Si and (b) $Ni_{81}Fe_{19}$ (111) in W/Py/TiN/Si thin films, confirming the epitaxial quality of the $Co_{60}Fe_{40}$ and $Ni_{81}Fe_{19}$ thin films grown on TiN buffered Si (100) substrates. STEM images for (c) W/$Co_{60}Fe_{40}$/TiN/Si, and (d) W/Py/TiN/Si thin films.

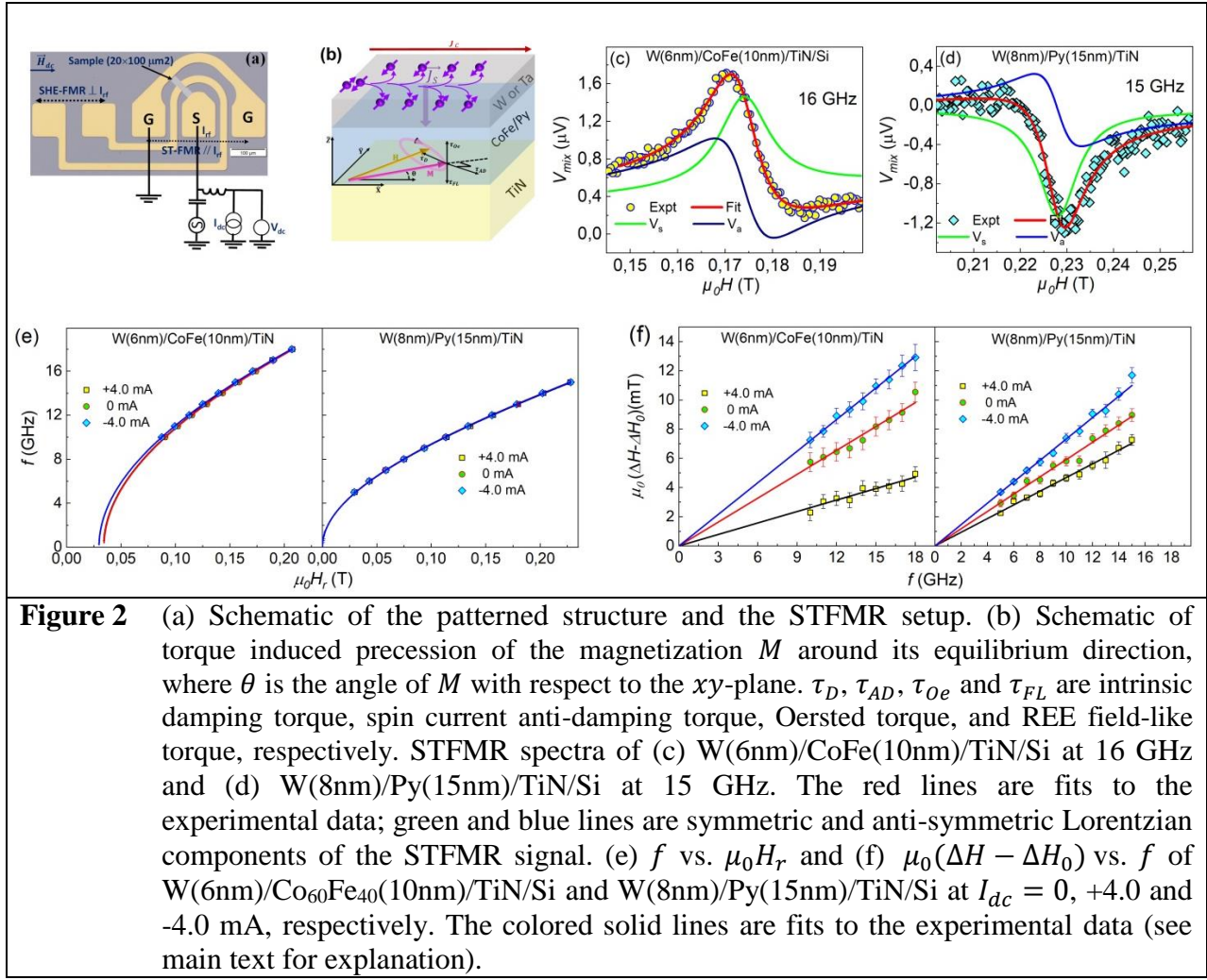

**Figure 2**  (a) Schematic of the patterned structure and the STFMR setup. (b) Schematic of torque induced precession of the magnetization $M$ around its equilibrium direction, where $\theta$ is the angle of $M$ with respect to the $xy$-plane. $\tau_D$, $\tau_{AD}$, $\tau_{Oe}$ and $\tau_{FL}$ are intrinsic damping torque, spin current anti-damping torque, Oersted torque, and REE field-like torque, respectively. STFMR spectra of (c) W(6nm)/CoFe(10nm)/TiN/Si at 16 GHz and (d) W(8nm)/Py(15nm)/TiN/Si at 15 GHz. The red lines are fits to the experimental data; green and blue lines are symmetric and anti-symmetric Lorentzian components of the STFMR signal. (e) $f$ vs. $\mu_0 H_r$ and (f) $\mu_0(\Delta H - \Delta H_0)$ vs. $f$ of W(6nm)/Co$_{60}$Fe$_{40}$(10nm)/TiN/Si and W(8nm)/Py(15nm)/TiN/Si at $I_{dc} = 0$, +4.0 and -4.0 mA, respectively. The colored solid lines are fits to the experimental data (see main text for explanation).

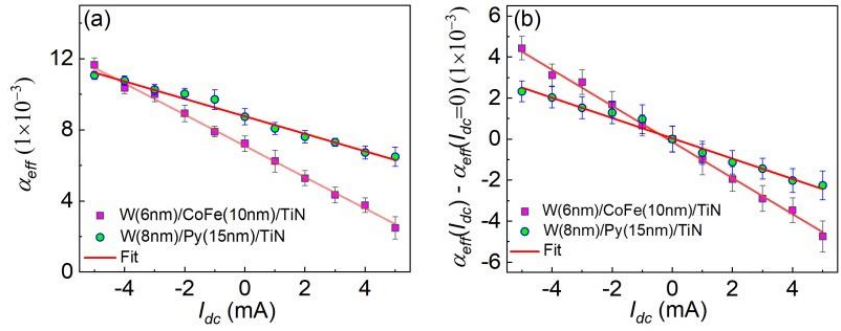

**Figure 3** (a) Effective damping parameter $\alpha_{eff}$ vs. $I_{dc}$ and. (b) $\alpha_{eff}(I_{dc}) - \alpha_{eff}(I_{dc} = 0)$ vs. $I_{dc}$ for positive applied fields for W(6nm)/CoFe(10nm)/TiN/Si and W(8nm)/Py(15nm)/TiN/Si. The solid lines are linear fits to the experimental data.

**Figure 4: Hall bar device measurements**

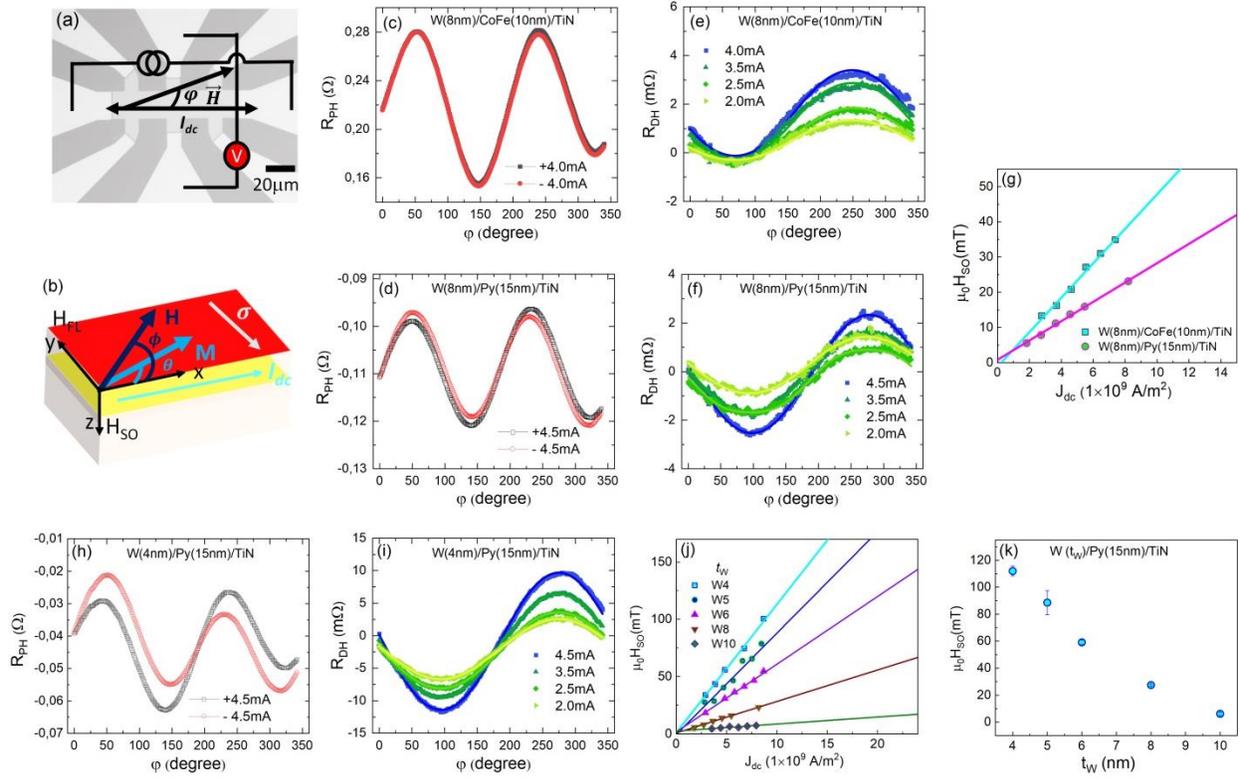

**Figure 4**  (a) Optical image of Hall-bar device with circuit measurement geometry. (b) In-plane magnetic field $\vec{H}$, magnetization $\vec{M}$ and fields generated by SOTs; anti-damping $\vec{H}_{SO}$ and field-like $\vec{H}_{FL}$. Angle dependent PHE resistance for (c) W(8nm)/CoFe(10nm)/TiN at $I_{dc} = \pm 4.0$ mA and (d) W(8nm)/Py(15nm)/TiN at $I_{dc} = \pm 4.5$ mA and corresponding $R_{DH}$ plots in (e) and (f) where solid lines are fits to Eq. (4). (g) $\mu_0 H_{SO}$ vs. $J_{dc}$ for both structures, where solid lines are linear fits to the data. (h) Angle dependent PHE resistance at $I_{dc} = \pm 4.5$ mA and corresponding $R_{DH}$ plots in (i) at different $I_{dc}$ for W(4nm)/Py(15nm)/TiN. Solid lines are fits to Eq. (4). (j) $\mu_0 H_{SO}$ vs. $J_{dc}$ for W(4-8nm)/Py(15nm)/TiN structures. Solid lines represent linear fits to the data. (k) $\mu_0 H_{SO}$ vs. $t_W$ for W($t_W$)/Py(15nm)/TiN structures at $J_{dc} = 1 \times 10^{10}$ A/m$^2$.

**Figure 5: Kerr microscopy images and results from micromagnetic simulations**

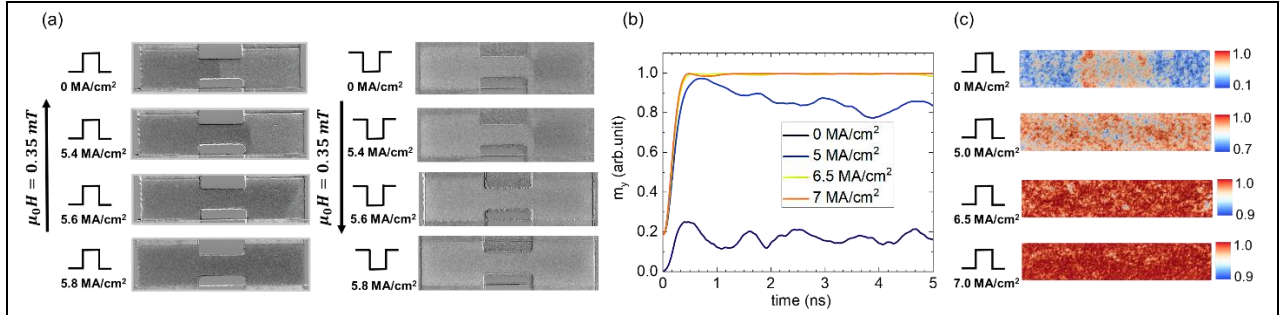

**Figure 5**     MOKE images of the SOT-driven magnetization switching by applying (a) a magnetic field of 0.35 mT along $+(-)y$-axis and a pulsed dc of increasing amplitude along $+(-)x$-axis for the W(8nm)/Py(15nm)/TiN structure. (b, c) Micromagnetic simulated magnetization ($y$-component) switching behavior of a $0.2 \times 1.0$ μm$^2$ structure modeled by applying a constant magnetic field along the $+y$-axis and a pulsed dc of increasing amplitude along the $+x$-axis for the W(8nm)/Py(15nm)/TiN structure. Changes in contrast visualizes the magnetization switching.